# A Reputation System for Market Security and Equity

Anton Kolonin,  Deborah Duong, Ben Goertzel, Cassio Pennachin, Matt Iklé , Nejc Znidar, Marco Argentieri

SingularityNET Foundation, Amsterdam, Netherlands
{anton, deborah,ben, cassio}@singularitynet.io

## Abstract

We simulate a reputation system in a market to optimize the balance between market security and market equity.  We introduce a method of using a reputation system that will stabilize the distribution of wealth in a market in a fair manner.  We also introduce metrics of a modified Gini that takes production quality into account, a way to use a weighted Pearson as a tool to optimize balance.

## 1   Introduction

Economists and politicians have noted a rising inequality in wealth. Barack Obama said that it had become "the defining issue of our time," back in 2013 [Parnass, 2013].  At the same time, we have seen online algorithms stand in for markets and determine who money flows to.  To contribute to study the relation between the growing inequality and online algorithms, we present preliminary example algorithms and metrics that explore what responsibility online algorithms that serve as artificial social proxies for social behavior might bear?

The online market is a different world than before, and these differences can contribute to the rise in inequality. The first exploitable difference online markets have that can lead to an unequitable market is anonymity.  Customer ratings and recommendations are often the only knowledge base a market has to match sellers with new customers, so that they may be "known" however the customers themselves are often "unknown". The first thing about online markets that prevent fair trade is people taking advantage of anonymity through scam reviews.  However, the defenses we have against the unfairness of scam reviews can introduce another kind of unfairness, one that comes from the fact that there is a tradeoff between safety from scam trades in markets and having all persons participate in the market in proportion to the quality of their offering.  Taken to extremes, to be entirely safe from scammers we could eliminate from the market all persons that had any chance of being a scammer, and only retain a few trusted sellers in each category.  However, this would keep low resource and new persons from entering the market.  On the other hand, the more chance we give to new sellers in the market to overcome initial bad reviews, the more opening we give to scammers.  A scammer can have artificial agents or paid co-conspirators give them fake good ratings, while the real ratings, which would be poor, could be mistaken for the ratings of those new to the business.  Economies of scale and ability to copy many goods for nothing make the matter worse - all the business goes to the very top.  In terms of machine learning and statistics, this can be thought of the tug between precision and recall, between being sure that when you declare an agent to be an honest agent , it is actually an honest agent,and being sure that you are giving every honest agent a chance to participate in the market. Except here the tradeoff is between security and equity. In order to assess how a market is doing in the tradeoff between security and equity, we have introduced an "F1" to tell us which side we are on, the weighted Pearson correlation coefficient of reputation scores.  We give an example of an algorithm that successfully strikes a balance between the two.

However, the accuracy of an algorithm can be affected by how the algorithm is used: accuracy in being able to accurately rank honest sellers is not enough to promote an equitable society.  For example in the nineties, when Google first used PageRank, it conceived of pages as having authority according to the number of links coming into them, that were themselves weighted by authority [Brin and Page 1998].  At that time it was a passive observer in authority rather than an active participant.  But the more people use Google's PageRank-like algorithm, the more Google becomes an active participant in the market.  It could be that Google as an active participant is less accurate than google as a passive participant because of feedback effects from Google's own authority.  For example, if everyone trusts google, and comes to know pages because they are first or second in googles ranking, and people link to links on their webpages because of what they have found there, then the position of the first few pages that the algorithm judges as best becomes even more solidified in a self fulfilling prophecy, and those that are not in the top few will get no business.  Such instabilities through feedback can cause the algorithm to lose accuracy. In terms of reinforcement learning, by presenting results in such a way that the user only uses the top one or two, the market suffers from too much exploitation and not enough exploration.  The businesses that are not in the very top ranks may even be better than those in the top ranks but remain inadequately explored because they are not accessible. We explore usages of reputation systems that stabilize the balance between exploitation and exploration, in particular, roulette wheel selection, enforced for example by offering discounts so that businesses ranked below the very top can still participate in the economy.

To measure the effectiveness of the roulette wheel in striking the balance between exploration and exploitation, we introduce a new metric for inequity based on the Gini coefficient, but which includes a measure of quality, defining inequity as wealth that strays from ones contribution in utility to the economy. We give an example of reputation rules and usage where greater equity results in greater utility, because of broader participation of businesses in the economy as well as better knowledge of what they can offer.

## 2 Methods

For these experiments we use the "liquid weighted rank" design of the reputation system per [Kolonin *et al.*, 2018] and its implementation according to [Kolonin *et al.*, 2019].

---

**Algorithm 1** Weighted Liquid Rank (simplified version)

**Inputs**:
1) Volume of rated transactions each with financial value of the purchased product or service and rating value evaluating quality of the product/service, covering specified period of time;
2) Reputation ranks for every participant at the end of the previous time period.
**Parameters**: List of parmeters, affecting computations - default value, logarithmic ratings, conservatism, decayed value, etc.
**Outputs**: Reputation ranks for every participant at the end of the previous time period.
1: **foreach of** *transactions* **do**
2:   **let** *rater_value* be rank of the rater at the end of previous period of default value
3:   **let** *rating_value* be rating supplied by   trasaction rater (consumer) to ratee (supplier)
4:   **let** *rating_weight* be financial value of the transaction of its logarithm, if logarithmic ratings parameter is set to true
5:   **sum** *rater_value\*rating_value\*rating_weight* for every ratee
6: **end foreach**
7: **do** normalization of the sum of the muliplications   per ratee to range *0.0-1.0*, get *differential_ranks*
8: **do** blending of the old_ranks known at the end of   previous peiod with differential_ranks based on  parameter of conservatism, so that *new_ranks = (old_ranks\*conservatism+N\*(1-differential_ranks))*, using decayed value if no rating are given to ratee   during the period
9: **do** normalization of *new_ranks* to range *0.0-1.0*
10: **return** *new_ranks*

---

We apply this algorithm to a marketplace of one thousand agents trading ten goods over six months. We simulated scam agents that give false ratings high ratings to a scam supplier, such that the market volume ratio of good agent trades and ratings to scam agent "trades" and false ratings was 50 and over, as one might see in a healthy market or perhaps in a market in which the goods that the agents scammed with were of lower cost (for example, if the raters were required to buy the good to be counted in the reputation system). We first use this reputation system on a market that has no overlap of consumers and suppliers, so that raters are not rated. This is typical of many consumer marketplaces designed for end users, such as the Amazon marketplace. We show that the reputation system shows both accuracy of rating honest agents and also of culling the scammers, and that this results in a lower loss to scam than if no reputation system were used. Next we apply the same reputation system to a system that has ninety percent overlap in the consumers and suppliers, such as we might see in a market that is not focused on the end user, such as ta B2B software market, so as to demonstrate the "liquid" part of the "weighted liquid rank." Here we compare two usages of the algorithm, first a "Winner Take All" usage in which consumers are encouraged by the presentation of the reputation scores to pick the highest scored suppliers, and the "Roulette Wheel" usage in which consumers are given discounts or otherwise presented with the results in a way that encourages them to choose suppliers in proportion to their reputation scores.

---

**Algorithm 2** Market Simulation

**Input**: Consumer and Supplier trade behaviors

**Output**: Metrics, Agent qualities, transactions, ranks

1: Assign Agents to Behaviors based on Normal Random Variates
2: Every day for 6 months:
     Each consumer makes shopping list

Agents drop past suppliers according to satisfaction

        If reputation system in use:
            If "winner take all" usage:
                Agents choose new suppliers with the highest
                    reputation score
            Else if "roulette wheel" usage:
                Agents choose new suppliers in proportion to
                    their reputation scores
            Else if "thresholded random" usage:
                Agents choose new suppliers randomly over a
                    reputation score threshold
                Agents make purchases and rate suppliers
        Else if reputation system not in use:
            If agents have no experience with suppliers
        Agents make purchases and rate suppliers
3: **Print** metrics

To capture the tradeoff between security and equity we offer the weighted Pearson correlation coefficient. The standard Pearson correlation coefficient between the "quality" of a sellers goods and his actual reputation score is used in the referenced work [Kolonin et al., 2019]. However, with the standard Pearson one can't tell if a value is because the flawed "security" or missed "equity". To contrast, the Pearson weighted towards the lowness of the reputation scores informs us how well the system is doing at security and the Pearson weighted towards the highness of the reputation scores tells how the system is doing on equity.

$$PCCW = \frac{cov(x,y,w)}{\sqrt{cov(x,x,w)*cov(y,y,w)}} \text{ where } cov(a,b,w) = \frac{\sum w(x-avg(x,w))(y-avg(y,w))}{\sum(w)}$$

We apply two social metrics : inequity, or how equal the society is, and utility, how satisfied the agents are with their purchases. An individual agent is treated with equity if it can engage in the economy in proportion to its talent. That is, if individual market volume/individual goodness are all somewhat equal. We use the Gini coefficient on individual market volume/individual goodness, so that, instead of measuring wealth, we let individual market volume stand in for wealth and additionally require that wealth should be proportional to the quality of an individual's goods, or "talent". By including talent, this metric predicts the price in a fair market, and posits that deviations of this price may arise from different knowledge of talent, as might originate in a biased reputation system. If this metric is high, the more unrelated trade is to talent.

The wealth of agent a, is $V_a = (V_{xa} + V_{ax})/2$, where $V_{xa}$ is market volume received by agent a and $V_{ax}$ is market volume spent by agent a. The equitable share of agent a is $W_a = V_a/R$, where R a reputation score between zero and one. We replace wealth value in the Gini coefficient [Gini, 1921] with this equitable share, where i is the sorted index, and N is the number of agents, as follows :

$$B = \frac{\sum_{a,i} \frac{sorted(Wa)(N-i)}{Rcea}}{N(\sum Wa)}$$

inequity = $(1 - \frac{1}{N} - 2B)$

The utility metric measures the satisfaction of agents with their purchases as the average rating given to purchases, regardless of the kind of purchase.

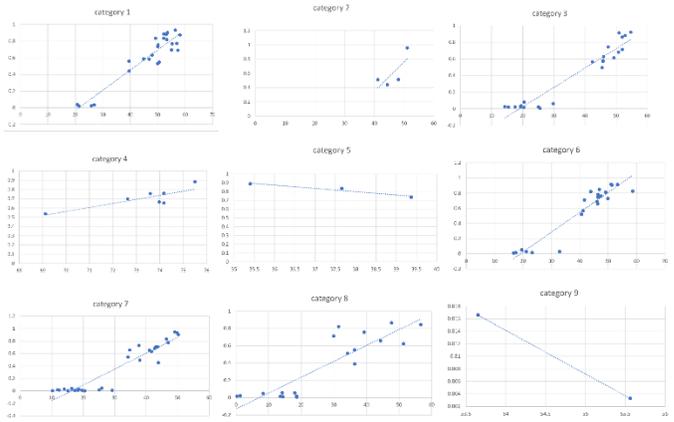

Figure 1. These charts from a single sample run show the Reputation system score along the X axis and the actual product quality on the Y axis, separated by good or service category. High Pearson by category coefficients in every category with over three samples indicate that a balance has been struck between securing the market from scammers and giving honest agents a fair ranking.

## 3 Results

The results of the first experiment, with no consumer and supplier overlap, are illustrated in figure 1. The average correlation between actual quality of products and the reputation system rating reached 0.95 for Pearson by good, with 0.92 weighted by honest agent and 0.96 weighted by scamming agent. This means that the high score of 0.95 was more attributable to the accuracy of reputation score more than it was attributable to accuracy of the scores of honest agents, although both are high.

In the second experiment for markets where consumer and supplier agents do overlap, such as in B2B markets, we confirmed that the Winner Take All usage of a reputation system results in both greater inequity and lesser utility than the

Roulette Wheel use of the reputation system. For our utility measure, Winner take all usages resulted in an average of 0.76 for all honest consumers, where Roulette wheel resulted in an average utility of .89, a difference significant at the 99% threshold. This means that in ratings from 0 to 1, the market as a whole was more satisfied when they choose suppliers in proportion to their reputation scores than when they chose the best reputation scores. At the same time, Roulette Wheel usage was far fairer, with wealth inequity scores an average of 0.26 while WTA usage had wealth inequity scores that averaged 0.43, a difference significant at the 99% threshold. This points to a a significant finding, demonstrating that fairness and inclusiveness also results in better products.

## 4 Discussion

These preliminary results, in two different types of online markets, show that a reputation system capable of striking a balance between security and equity can be used to increase both equity in distribution of business opportunity at the same time as it can achieve better products for customers. As long as it can both detect scams and rank honest agents fairly, users can be offered incentives to purchase the goods and services of agents in proportion to their score. Such a usage of a reputation system will not only ensure that businesses get a fair chance to participate in the economy, but also the products made in such a decentralized market will result in greater satisfaction to consumers. If we view the economy as a reinforcement learning system, where the growing inequality can be interpreted as the vice of too much exploitation and not enough exploration, perhaps advances in AI can help us to ensure a fairer society.

## References

[Brin and Page, 1998] Brin S. and Page L. *The Anatomy of a Large Scale Hypertextual Web Search Engine, Computer Network and ISDN Systems, Vol. 30, Issue 1-7, pp. 107-117, 1998.*

[Duong, 1996] SISTER: A Symbolic Interactionist Simulation of Trade and Emergent Roles. JASSS Fall 1996.

[Kolonin et al., 2018] Kolonin A., Goertzel D., Duong D., Ikle M. A Reputation System for Artificial Societies. arXiv:1806.07342 [cs.AI], 2018.

[Mesa, 2016] Mesa Project Team, George Mason 2016. https://mesa.readthedocs.io/en/master/packages.html#references

[Parnass, 2013] Sarah Parnass. *Full transcript: President Obama's December 4 remarks on the economy.* The Washington Post, Washington DC December 14, 2013.